\documentclass[conference]{IEEEtran}
\usepackage{graphicx}
\usepackage{amssymb}

\begin{document}


\title{Self-Organized Authentication Architecture for Mobile Ad-hoc Networks}



\author{\IEEEauthorblockN{P. Caballero-Gil\IEEEauthorrefmark{1}, C. Caballero-Gil\IEEEauthorrefmark{2}, J. Molina-Gil \IEEEauthorrefmark{2}, C. Hern\'andez-Goya\IEEEauthorrefmark{1}}

\IEEEauthorblockA{\IEEEauthorrefmark{1}Dept. Statistics,
Operations Research and Computing. University of La Laguna. 38271
La Laguna. Tenerife.\\ Spain. E-mail: {pcaballe,mchgoya}@ull.es }

\IEEEauthorblockA{\IEEEauthorrefmark{2}Dept. Informatics and
Systems. University of Las Palmas de Gran Canaria. 35017 Las
Palmas de Gran Canaria.\\ Gran Canaria.  Spain. E-mail:
{candido.caballero,jezabelmiriam}@gmail.com }

}

\maketitle

\begin{abstract}
This work proposes a new architecture, called Global Authentication Scheme for Mobile Ad-hoc Networks (GASMAN), for fully distributed and self-organized authentication. In this paper apart from describing  all the GASMAN components, special emphasis is placed on proving that it fulfils every requirement of a secure distributed authentication scheme, including limited physical protection of broadcast medium, frequent route changes caused by mobility, lack of structured hierarchy, etc. Furthermore, an extensive analysis through simulation experiments in different scenarios is described and discussed.

\end{abstract}

\section{Introduction}
\label{Sec:Intro}
\footnotetext{Work developed in the frame of the project TIN2008-02236/TSI supported by the Spanish Ministry of Science and Innovation and FEDER Funds.\\
Proceedings of the 6th International Symposium on Modeling and Optimization in Mobile, Ad Hoc, and Wireless Networks, WiOpt , (April 2008).}

Network security consists of several services such as authentication, confidentiality, integrity, non-repudiation, and access control. Among these facilities, authentication, which ensures the true identities of nodes, is the most fundamental one because other services depend fully on the sure authentication of communication entities.

Mobile Ad-hoc NETworks (MANETs) are autonomous networks formed by mobile nodes that are free to move at will. These networks have received increasing interest in the last years, partly owing to their potential applicability to myriad applications, ranging from small, static networks that are constrained by power sources, to large-scale, mobile and highly dynamic networks.
Conventional wired networks mainly use a globally trusted
Certificate Authority (CA) for solving the authentication problem.
However, authentication in MANETs is undoubtedly much more
difficult to solve due to several reasons such as the absence of a
fixed infrastructure and centralized management, the dynamic
nature of the nodes, the limited wireless range of nodes, the
dynamic topology, frequent link failures and possible transmission
errors \cite{ATEDQ05} \cite{Wei04}. Also, since all the nodes must
collaborate to forward data, the wireless channel is prone to
active and passive attacks by malicious nodes, such as Denial of
Service (DoS), eavesdropping, spoofing, etc.

One of the most elementary approaches to authentication in MANETs
found in the bibliography uses a Trusted Third Party (TTP) to
guarantee the validity of all nodes identities so that every node
who wants to join the network has to get a certificate from the
TTP. A second identification paradigm that has been used in
wireless ad-hoc networks is the notion of  chain of trust
\cite{HBC01}. A third typical solution is location-limited
authentication, which is based on the fact that most ad-hoc
networks exist in small areas and physical authentication may be
carried out between nodes that are close to each other. The
special nature of ad-hoc networks, where most applications are
collaborative and group-based, suggests that such traditional
approaches to node identification may not be always appropriate.
As an example of this fact, we have that different authentication
protocols have been recently proposed for ad-hoc networks but all
of them have some drawback. For instance, the work \cite{HJYSCL}
is based on the RSA signature, which conducts to the problem of
public key certification. Another recent example is the paper
\cite{STY05}, which provides a solution that works well just for
short-lived MANETs. Thus, the design of a new
scheme that fulfils all the requirements for this type of networks
continues being considered an open question.

Here we propose a new architecture for authentication in ad-hoc
networks called Global Authentication Scheme for Mobile Ad-hoc
Networks (GASMAN), which is based on the established cryptographic
paradigm of Zero-Knowledge Proofs (ZKPs). Since the information
sent while its execution does not convey any secret related to the
authentication process, ZKPs provide an elegant and fault-tolerant
solution to node authentication in MANETs. Furthermore, the GASMAN
has other three beneficial properties. Firstly, it has scalability
because centralized elements, such as CA servers, are not
required. Secondly, availability is guaranteed through the
insertion, deletion and access control procedures. Finally, our
architecture assures strong authentication to any legitimate node
willing to join the network by using the zero-knowledge proof
implemented in the access control algorithm. The GASMAN algorithms
jointly with mobility help to reduce the time necessary for nodes
to join and access the network in a timely manner. Summing up, the
main features of the proposed protocol are the adaptation to the
varying topology of the network, the open availability of
broadcast transmissions and the strong access control.

Up to now, very few
publications have  mentioned the proposal of authentication
systems for ad-hoc networks using ZKPs \cite{ARMF06}
\cite{WZK05}, and none of them has dealt with the related problem
of topology changes in the network. A recent ZKP-based
hierarchical proposal for MANETs related with the one proposed
here was described in \cite{CH06}, where two different security
levels were defined through the use of a hard-on-average graph
problem, and no topology changes were considered.

This work is organized as follows. The following section provides
an overview of the proposal, some general aspects of the proposal and the notation used in its description. Specific details about the five principal elements of the architecture, i.e.,  network initialization, node insertion, access control, proofs of life and node deletion are gathered in Section \ref{ElementsDescription}. The assumptions required by GASMAN and an analysis of its security are
commented in Section \ref{AssumptionsasnSecurity}. A performance analysis developed under NS-2 is provided in
Section \ref{PerformanceAnalysis}. Finally, some conclusions and open questions complete
the paper.

\section{Basics and Notation}

\label{BasicsandNotation}

The proposed protocol was designed as an authentication scheme for
group membership because when a node wants to be part of the
network, it has to be previously authorized by a legitimate node.
According to the authors of \cite{MAH00}, in any group member
authentication protocol it is necessary to provide robust methods
to insert and to delete nodes, as well as to allow the access only
for legitimate members of the group. For that reason, not only the
ZKP used for access control is described, but also the
update procedures associated to insertions and deletions are
carefully defined. For instance, the procedure to delete nodes is only initiated once a node has been
disconnected of the network for too long. The period of time after which the node is deleted is a parameter ($T$) of the system here presented.

The access control described below is based on the general scheme
of Zero-Knowledge Proof introduced in \cite{CH01}, when using the
Hamiltonian Cycle Problem (HCP). A Hamiltonian cycle in a graph is
a cycle that visits each vertex exactly once and returns to the
starting vertex. Determining whether such cycles exist in a graph
defines the Hamiltonian Cycle Problem, which is NP-complete. Such
a problem was chosen for our design mainly due to the low cost of
the operations associated to the update of a solution. This is an
important characteristic since in a highly dynamic setting such as
MANETs these operations will be developed frequently. Anyway,
there should be pointed out that similar schemes based on
different NP-complete graph problems might be described. The only
feature demanded to the problems chosen is that the solutions may
be easily updated when small changes occur in the network. This is
just the case of the Vertex Cover, Independent Set or Clique
Problems, for instance.

One of the key points to assure the correct operation of GASMAN is the use of a chat application through broadcast that
makes it possible for legitimate on-line nodes to send a message
to all on-line users. Such an application allows
publishing all the information associated to the update of the
network. Although secrecy is not necessary for chat messages, because they are not more than gibberish for illegitimate nodes, it is required that only the on-line legitimate nodes execute the chat application.

The information received through the chat application during an
interval of time must be stored by each on-line node in a FIFO
queue. Such data should be stored by each on-line node, allowing in this way the updating of
the authentication information not only to it but also to all the
off-line legitimate nodes whose access will be granted. Such a period is an essential
parameter in the system because it states both the maximum off-line time allowed
for any legitimate node, and the frequency of broadcasting the proofs
of life. Consequently, such a parameter should be previously
agreed among all the legitimate nodes of the network.

A generic life-cycle of a MANET has three major phases that are
described below (see figure 1
):
\begin{description}
    \item [\emph{Initialization}:]  $\\$ Each initial member in the original network will be securely provided,
either off-line or on-line, with a secret piece of information. The knowledge of the secret network key will be used during access
control in order to prove the node's eligibility for accessing to the protected resources or to offer service to the network. After completing this stage, the legitimate nodes are ready to actively participate in the network.
    \item [\emph{Access Control}:] $\\$The access control process allows a legitimate node to prove its network membership to an on-line node. These legitimate nodes must demonstrate knowledge of the secret network key by using a challenge-response scheme.
    \item [\emph{On-line Session:}] $\\$Once the legitimate node reaches an on-line state in the network, it gets full access to the protected resources such as the chat application. At the same time, it may offer services such as the insertion of new nodes. There should be taken into account that the secret network key will be updated according to the network evolution. Hence, if a node is off-line for too long, its secret key will expire. In such a case, the legitimate node would have to be re-inserted by an on-line legitimate node.

\end{description}

\begin{figure*}[htb]
  \centering
     \includegraphics[scale=0.35]{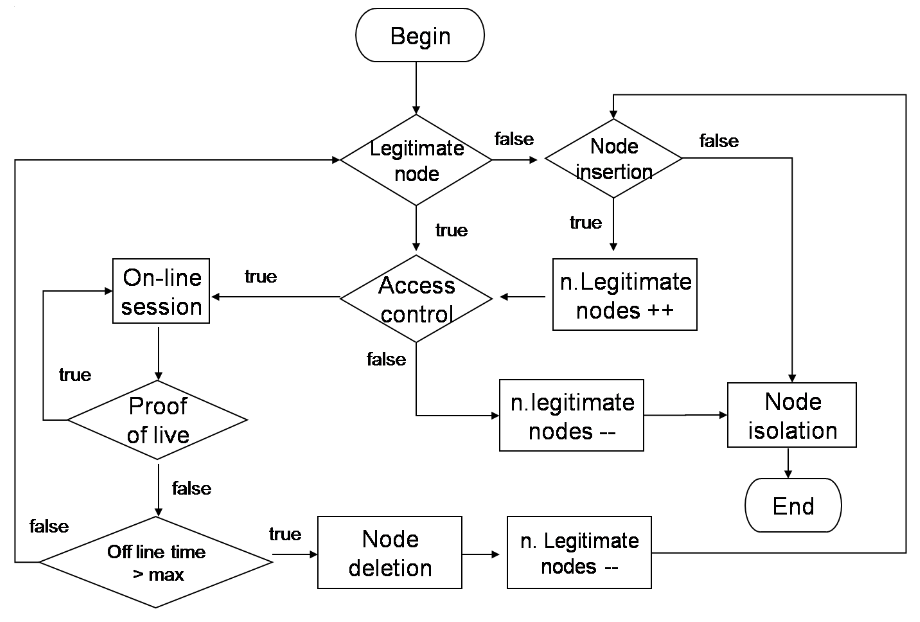} 
   \caption{Node Life-Cycle}
   \label{fig2}
\end{figure*}

Since in our proposal the secrecy of the network key is provided
by the difficulty of the HCP, the number of on-line legitimate
nodes is a crucial parameter. In consequence, as soon as the
number of on-line legitimate nodes becomes too small (when
comparing it with certain threshold parameter), the network
termination is carried out and therefore, the life-cycle of the
network ends.

Probably, the most remarkable aspect in our proposal is that no meaningful information may be stolen even if an adversary is able to reads the whole information published through the chat application, or even if it eavesdrops the information exchanged between a legitimate prover and verifier at the time of executing the access control protocol.

In the following, the basic notation used throughout the proposal is explained.

\begin{itemize}
\item $G_t=(V_t,E_t)$ denotes the undirected graph used at stage $t$ of the network
life-cycle.

\item  $v_i \in V_t$ represents both a vertex of the graph and a legitimate node.

\item $n=\left|V_{t}\right|$ is the order of $G_t$, which coincides with the
number of legitimate nodes.

\item $N_{G_{t}}(v_i)$ denotes the neighbours of node $v_i$ in the graph $G_t$.

\item ${\Pi}(V_t)$ represents a random permutation over the
vertex set $V_t$

\item ${\Pi}(G_t)$ denotes the graph isomorphic to $G_t$ built after applying permutation ${\Pi}(V_t)$.

\item $c \in_r C$ indicates that an element $c$ is chosen at random with uniform distribution from a set $C$.

\item $HC_t$ designates the Hamiltonian cycle used at stage $t$.

\item ${\Pi}(HC_t)$ represents the Hamiltonian cycle $HC_t$ in the graph
${\Pi}(G_t)$.

\item $N_{HC_t}(v_i)$ denotes the neighbours of node $v_i$ in the Hamiltonian
cycle $HC_t$.

\item $S$ and $A$ stand for the supplicant and the
authenticator, respectively. This notation is used both while an insertion phase and the
execution of a ZKP-based access control are carried out.

\item $S \rightleftharpoons A$ symbolizes when node $S$ contacts $A$.

\item $A \leftrightarrow S: information $  means that $ A $ and $S$ agree on $information$

\item $A \stackrel{s}{\rightarrow} S: information $  means that $ A $ sends $information$
to $S$ through a secure channel.

\item $A \stackrel{o}{\rightarrow} S: information $  means that $ A $ sends $information$
to $S$ through an open channel.

\item $A \stackrel{b}{\rightarrow} network: information $ represents when $
A $ broadcasts $information$ to all on-line legitimate nodes.

\item $A \stackrel{b}{\leftrightarrow} network:
information $ represents a two-step procedure where $ A $
broadcasts $information$ to all on-line legitimate nodes of the
network, and receives their answers.

\item $h$ stands for a public hash function.

\item $T$ denotes the
threshold period that a legitimate nodes may be off-line without been excluded of the network.

\end{itemize}

\section{GASMAN description}
\label{ElementsDescription}
This section contains the description of the procedures that form part of the GASMAN architecture, including all the specific details
about network initialization, node insertion, access control,
proofs of life and node deletion.
\subsection{Network Initialization}

The proposed protocol requires the
definition of an initialization phase where the secret information
associated to the process of identification is generated and
distributed within the initial network. This initialization phase
consists in the definition of the graph used for the development
of the protocol. Such a graph should be generated with the participation of all the original members of the
network. Furthermore, the initialization phase also implies the shared generation by the initial legitimate members of the network of a solution to the HCP in such a graph.

In our proposal, as in trust graphs \cite{JB}, the graph vertexes set corresponds to the set of nodes in the actual network during its whole life-cycle. Consequently, the initialization process starts from a set $V_{0}$ of $n$
vertexes corresponding to the nodes of the initial network.
Hence, each vertex sub-index may be used as ID
(IDentification) for the corresponding node. The first step of the
initialization process consists of generating cooperatively and secretly
a random permutation $\Pi$ of such a set. Once this generation is
completed, each legitimate node should know a Hamiltonian cycle
$HC_{0}$ corresponding exactly to such a permutation. Finally, the
partial graph formed by the edges corresponding to such a
Hamiltonian cycle $HC_0$, is completed by adding $n$ groups of
$\frac{2m}{n}$ edges, producing the initial edge set $E_0$. Here, $m$ stands for the number of edges that the initial graph will have after the initialization stage. Each one of these $n$ groups of edges will be generated by $v_i$, $i=1,2,...,n$ according to the following restrictions: they must have $v_i$ as one of its vertexes,
while the other one will be randomly generated. Note
that the size $\frac{2m}{n}$ of those edge subsets must
be large enough so that the size of the resulting edge set
$\left|E_{0}\right| = m$ guarantees the difficulty of the HCP in
the graph $G_0$.

\begin{description}[Initialization Algorithm]
\item [Input:] $V_{0}$, with $\left|V_{0}\right| = n$.

\item [1.]The  $ n $ nodes of the network generate cooperatively, secretly and randomly the cycle $HC_{0}=\Pi\left(V_{0}\right)$.

\item [2.] $\forall v_i \in V_{0},\ v_i$ builds the set $$N_{G_{0}}(i) =\left\{   \{v_j \in_r V_{0}\} \cup N_{HC_0}(i)\right\}$$ with $\left|N_{G_{0}}(i)\right| = \frac{2m}{n}$.

\item [3.] $\forall v_i \in V_{0}: v_i \stackrel{b}{\rightarrow} network: N_{G_{0}}(i) $.

\item [4.] $\forall v_i \in V_{0}: v_i$ merges: $$ E_{0}= \bigcup _{i=1,2,...,n} \left\{ (v_i, v_j) : v_j \in N_{G_{0}}(i) \right \}.$$

 \item [Output: ] $\ G_{0} = (V_{0}, E_{0})$, with $\left|E_{0}\right| = m $.
\end{description}

Once the creation of the initial instance of the problem has
been carried out through the contribution of all the nodes
of the network, each node will know a Hamiltonian
cycle in the resulting $\frac{2m}{n}$-regular graph. From then on,
each time a new user $S$ wants to become a member of the network,
it has to contact a legitimate member $A$ in order to follow the
insertion procedure explained in the following section.

\subsection{Node Insertion}

Let us suppose that we are at  stage $t$ of the network life-cycle
when a user $S$ contacts a legitimate member $A$ of the network to
become a member of the network. Once $S$ has convinced $A$ to
accept its membership in, the first step that $A$ should
carry out is to assign $S$ the lowest vertex number $v_i$ not assigned
so fat in the vertex set $V_t$. Afterwards, $A$ should
broadcast such an assignment to all on-line legitimate nodes in order to prevent another simultaneous insertion with the same identifier. If $A$ receives
less than $n/2$ answers to the previous message, she stops the insertion procedure because
the number of nodes that are aware of the insertion is not large
enough. Otherwise, $A$ develops the corresponding update of the
secret Hamiltonian cycle $HC_t$ by selecting at random two
neighbour vertexes $v_j$ and $v_k$ in order to insert the new node
$v_i$ between them. Additionally, $A$ chooses at random a subset of $\frac{2m}{n} -2$
nodes in $V_t$ such that none of them is its neighbour in $HC_t$. Finally, $A$ broadcasts the set of neighbours $N_{G_{t+1}}(v_i)$ of $S$ in
the new graph $G_{t+1}$ .

Each time a node receives a graph update, it should secretly modify the corresponding Hamiltonian cycle. In order to achieve it, it uses the information provided to identify the unique position (according to the new edge set $E_{t+1}$) in the cycle where the new node can be inserted. In this way, it will be able to easily update the secret network key by simply inserting the vertex $v_i$ between the vertexes $v_j$ and $v_k$. At the same time, the
authenticator node $A$ must send the supplicant node $S$ both the graph $G_{t+1}$ (deploying an open channel), and  the Hamiltonian cycle $HC_{t+1}$ (through a secure channel).

\begin{description}[Insertion Algorithm]

    \item  [Input:] At stage $t$ a supplicant node $S$ wants to become a member
of the network.
    \item [1.]$S\rightleftharpoons A$.
    \item [2.] Node $S$ convinces node $A$ to accept its membership in the network.
    \item [2.] $A$ assigns $S$ the identifier $v_i$ such that $i = min \{l: v_l \not \in V_t \}$
    \item [3.]$ A \stackrel{b}{\leftrightarrow} network: v_i$
    \item \begin {itemize}
      \item [3.1] If $A$ receives less than $n/2$ answers, she stops the insertion procedure.
      \item [3.2]Otherwise:
    \begin{enumerate}
     \item [3.2.1]$A$ chooses:\\
            $(v_j,v_k): v_j\in _r V_t,v_k \in _r N_{CH_t}(v_j)$
     \item [3.2.2]$A$ chooses at random:\\
     $$N_{G_{t+1}}(v_i)= \{v_j, v_k\} \cup \{ w_1,w_2,..., w_{\frac{2m}{n} -2}\}$$  such that
    $ N_{G_{t+1}}(v_i) \subseteq V_t \wedge \forall w_{l_1},w_{l_2}: w_{l_1} \not \in N_{CH_t}(w_{l_2})\}$
    \item [3.2.3] $ A \stackrel{b}{\rightarrow} network: N_{G_{t+1}}(v_i)$
    \item [3.2.4]Each on-line node updates $G_t$ by defining $V_{t+1}=V_t \cup \{v_i\}$, $E_{t+1}=E_t \cup N_{G_{t+1}}(v_i)$ and
     $HC_{t+1}= HC_t \setminus \{(v_j, v_k)\} \cup \{(v_j,v_i) \cup (v_i, v_k)\}$
    \item [3.2.5] $A \stackrel{o}{\rightarrow} v_i: G_{t+1}$
    \item [3.2.6] $A \stackrel{s}{\rightarrow} v_i: HC_{t+1}$
\end{enumerate}
\end {itemize}

\item [Output:] $\ $ The supplicant node $S$ becomes a legitimate member of the
network.
\end{description}

\subsection{Access Control}
\label{Sub:AcCont}

If a legitimate node $S$ has been off-line
or out-of-coverage from stage $t$ and wants to re-enter into the
network at stage $r$, its first step should be to contact a
legitimate on-line member $A$. Afterwards, $A$ should check
whether the period $S$ has been off-line is not greater than $T$. In
this case, $S$ has to be authenticated by $A$ through a ZKP based on its
knowledge of the secret solution $HC_t$ on the graph $G_t$.

The aforementioned ZKP begins with the agreement between $A$ and
$S$ on the number of iterations $l$ to execute. From there on, in
each iteration, $S$ will choose a random permutation
${\Pi}_j(V_t)$ on the vertex set that will be used to build a
graph ${\Pi}(G_t)$ isomorphic to $G_t$. The hash value of both the
permutation $h({\Pi}_j(V_t))$ and the Hamiltonian cycle in the
graph  $h({\Pi}_j(HC_t))$ are then sent to $A$. When this
information is received by $A$, it chooses a bit $b_j$ at random
($b_j \in_r \left\{0,1\right \}) $. Depending on the selected
value, $S$ will provide $A$ with the image  of the Hamiltonian
cycle through the isomorphism, or with the specific definition of
the isomorphism. In the verification phase, $A$ will check that
the received information was correctly built.

Once the authentication of supplicant $S$ has been successfully
carried out, the authenticator $A$ gives him the necessary
information to have full access to the protected resources such as
the chat application, for example.

\begin{description} [Access Control Algorithm]

\item  [Input:] At stage $r$ a supplicant node $S$ that has been off-line
since stage $t$ wants to re-enter into the network.
    \item [1.] $S\rightleftharpoons A$
    \item [2.] $S \stackrel{o}{\rightarrow} A: G_t$
    \item [3.] $A$ checks whether $r - t \leq T$
      \begin {itemize}
      \item [4.] if $r - t > T$ then $S$ is not authenticated
      \item [5.] otherwise:
      \begin{itemize}
        \item $A \leftrightarrow S: l $
        \item for $ j = 1,2,\cdots,l$
        \begin{enumerate}
             \item [5.1] $S$  chooses ${\Pi}_j(V_t)$ and builds  ${\Pi}_j(G_t)$ and ${\Pi}_j(HC_t)$, the graph
             isomorphic to $G_t$ and the corresponding Hamiltonian cycle, respectively.
             \item [5.2] $S \stackrel{o}{\rightarrow} A: \{ h({\Pi}_j(V_t)), h({\Pi}_j(HC_t)) \}$
             \item [5.3] $A$ chooses  the challenge $b_j \in_r \left\{0,1\right \} $
             \item [5.4] $A \stackrel{o}{\rightarrow} S: b_j$
             \begin{enumerate}
                    \item [5.4.1] If $b_j=0$ then $S \stackrel{o}{\rightarrow} A: \{ {\Pi}_j(G_t), {\Pi}_j(HC_t) \} $
                    \item [5.4.2] If $b_j=1$ then $S \stackrel{o}{\rightarrow} A: {\Pi}_j$
             \end{enumerate}
             \item [5.5] $A$ verifies that
             \begin{enumerate}
                \item ${\Pi}_j(HC_t))$ is a valid Hamiltonian cycle in ${\Pi}_j(G_t)$, if $b_j=0$
                \item the hash function $h$ applied on ${\Pi}_j(G_t)$ coincides with $h({\Pi}_j(G_t))$, if $b_j=1$
             \end{enumerate}
        \end{enumerate}
        \item if $\exists  j \in \left\{1,2,\ldots,l\right\}$ such that the verification is negative, then $S$ is isolated.
        \item otherwise $A \stackrel{s}{\rightarrow} S:$ the necessary information to have full access
            to protected resources of the network.
    \end{itemize}
    \end{itemize}
\item [Output:] $\ $ Node $S$ is connected on-line to the network.
\end{description}

\subsection{Proofs of Life}

All on-line legitimate nodes have to confirm their presence in an active way. Such a confirmation is carried out every period of time $T$. It consists in broadcasting a message (proof-of-life) to all on-line legitimate nodes.

If some insertion happens during such a period, a proof of life of every on-line legitimate node will be distributed together with the information necessary for the insertion procedure. Otherwise, only the proof of life is required. During such a broadcast every node adds its own proof of life to the broadcast. In this way, when the broadcast reaches the last node, a broadcast back starts containing the proofs of life of all on-line legitimate nodes.

\begin{description} [Proof-of-Life Algorithm]

\item [Input:] At stage $t$ node $A$ is an on-line legitimate node of the
network.
    \item [1.] $A$ initializes its $clock=0$ just after its last proof of
    life.
    \item [2.] if $clock > T$ then
    \begin{enumerate}
        [2.1]\item $ A \stackrel{b}{\leftrightarrow} network: A's$ $ proof$ $of$ $life$
     \begin {itemize}
     \item [2.1.1] If $A$ receives less than $n/2$ proofs of life as answers to her broadcast, she stops her proof of life and puts back her clock.
      \item [2.1.2] Otherwise: $ A \stackrel{b}{\rightarrow} network: Received$ $proofs$ $of$ $life$
\end {itemize}
    \end{enumerate}

    \item [Output:] $\ $ At stage $t+1$ node $A$ continues being an on-line
legitimate node of the network of the network.
\end{description}

\subsection{Node Deletion}

The deletion procedure is mainly based on the confirmation of the
active presence of on-line legitimate nodes through their proofs
of life. Each node should update its stored graph by deleting all
those nodes that have not sent any proof of life after a period
$T$. This fact implies that each node that has not proven its presence
will be deleted from the network, as well as from the Hamiltonian cycle.

Node deletions are explicitly communicated to all on-line
legitimate nodes in the second step of broadcasts of proofs of
life. This way to proceed allows any node that is off-line in that moment will
be able to update its stored graph as soon as it gets access to
the network.

\begin{description}[Deletion Algorithm]

\item [Input:] At stage $t$, a node $v_i$ is an off-line legitimate node of the network.
    \item [1.] $A$ initializes her $clock=0$.
    \item [2.] if $clock > T$ then
    \begin{enumerate}
    \item [2.1] $\forall v_i \in V_t$: $A$ checks $v_i$'s proof of life in $A$'s FIFO queue.
    \item [2.2] $A$ updates $V_{t+1}=V_t \setminus \{v_i\in V_t$ with no proof $\}$.
    \item [2.3] $A$ updates $E_{t+1}=E_t \setminus \{ (v_i, v_j): v_i \in V_t$
    with no proof, $v_j \in N_{G_t(v_i)} \} \cup \{(v_j, v_k): v_j, v_k \in N_{HC_t(v_i)} \}$.
    \item [2.4] $A$ updates $HC_{t+1}=HC_t \setminus \{(v_j, v_i), (v_i, v_k)\} \cup (v_j,v_k): v_i \in V_t$
        with no proof, $v_j, v_k \in N_{HC_t(v_i)}$
    \end{enumerate}
    \item [3.]  If $A$ started the broadcast used for the $v_i$'s deletion, $A$ adds this information
    to the second step of the proof-of-life broadcast:  $A \stackrel{b}{\rightarrow} network: $ $v_i$ is deleted.
\item [Output:] $\ $ At stage $t+1$ the node $v_i$ has been deleted both from the network and from the graph.
\end{description}

This procedure guarantees a limited growth of the graph that
is used in authentication, and at the same time, allows that
always the legitimate nodes set corresponds exactly to the
vertexes in that graph. Apart from this, it is remarkable the fact
that thanks to this procedure  the recovery of legitimate members
of the network that have been disconnected momentarily is possible.

\section{Assumptions and Security Analysis}
\label{AssumptionsasnSecurity}
This proposal initially assumes the ideal environment where all
legitimate nodes are honest and where no adversary may compromise
a legitimate node of the network in order to read its secret
stored information. Such assumptions are well suited as a basic
model in order to decide under which circumstances the GASMAN is applicable to MANETs. For instance, a possible adaptation of the proposal in order to avoid those
hypothesis could be defining a threshold scheme to be used in every step of the GASMAN, so that every proof of life, insertion, access control or deletion operation should be done by a coalition of on-line nodes. Then, a dishonest node would not affect
the correct operation of the network.

It is clear that the proposal inherits some problems of the distributed trust model such as the important necessity that legitimate nodes cooperate. Consequently, it is advisable to include a scheme to stimulate node cooperation.

Finally, another requirement of the GASMAN is the establishment of a secure channel for the insertion procedure. However, that aspect may be easily fulfilled thanks to the fact
that most wireless devices are able to communicate with each other via Bluetooth wireless technology.

With respect to possible attacks and due to the lack of a centralized
structure, it is natural that possible DOS (Denial Of Service)
attacks have as their main objective the chat application. In
order to protect the GASMAN against this threat it must be assured
that chat messages, although are publicly readable, may be only
sent by legitimate on-line members of the network. Another
important aspect related to the use of the chat application is the
necessary synchronization of the on-line nodes, so a common
network clock is necessary.  This requirement has been implemented
during simulations through the chat application.

MANETs are in general vulnerable to different threats such as
identity theft (spoofing) and the man-in-the-middle attack. Such
attacks are difficult to prevent in environments where membership
and network structure are dynamic and the presence of central
directories cannot be assumed. However, our proposal is resistant
to spoofing attacks because access control is granted through a ZKP. It implies
that any information published through the chat application or sent openly during the execution of access control mechanism becomes useless.

On the other hand, the goal of the man-in-the-middle
attack is  either to change a sent message or to gain some useful
information by one of the intermediate nodes. Again, the use of
ZKPs in our protocol implies that reading any transferred
information does not reveal any useful information about the
secret, so changing the message is not possible since only
legitimate nodes whose access has been allowed can use the chat
application.

Another active attack that might be especially dangerous in MANETs
is the so-called Sybil attack. It happens when a node tries to get
and use multiple identities. The most extreme case of this type of
attacks is the establishment of a false centralized authority who
states the identities of legitimate members. However, this
specific attack is not possible against our scheme due to its
distributed nature. In the GASMAN, the responsibility of
controlling general Sybil attacks will be shared among all the
on-line nodes. If an authenticator  node detects that a supplicant
node is trying to get access to the network by using an
 ID that is yet being used on-line, such access control must be denied and the corresponding node must be isolated. The same happens when any on-line node detects that
an authenticator node is trying to insert a new member into the
network with a new ID, and such a node has yet assigned as a vertex
ID. Again, such insertion must be denied and the corresponding
supplicant node must be isolated. Anyway, if a Sybil attacker
enters the network, any of its neighbours will detect it as soon
as it sends proofs of life for different vertexes ID.

\section{Performance Analysis}
\label{PerformanceAnalysis}
We now analyze the efficiency of the proposal both from the energy
consumption and from computational complexity points of view. We
consider the energy consumption which is the result of
transmissions of data and processor activities due to
authentication tasks. In the proposal there are two phases when
computational overhead is more significant: the ZKP-based access
control and the periodic checking of stored elements in the FIFO queue. A
reduction on the number of rounds of ZKP has a direct effect on
the total exchanged messages size in insertions, but a trade-off
should be maintained between protocols robustness and performance.
Indeed, regarding total data transmission over wireless links, the
ZKPs take less than 10\% in a usual situation.

The dominant time-consuming jobs are the periodic proofs of life, which accounts for around 90\% of the
total exchanged message size in many cases. However, we found
that these compulsory proofs of life imply an incentive technique
for stimulating cooperation in authentication tasks. This is due
to the fact that nodes that are broadcasters of deletion queries or
authenticators in insertions or access controls  are exempted from
their obligation to broadcast their proofs of life.

In order to reduce  data communication cost of the protocol, an
increase on the threshold period $T$ might be an option, but again
an acceptable balance should be kept. According to our
experiments, $T$ should depend directly on the number of
legitimate and/or on-line nodes in order to prevent a possible
bandwidth overhead in large networks.

For the performance analysis of the proposal we used the Network 
Simulator NS-2 with the DSR routing protocol.  We created several Tcl
based NS-2 scripts in order to produce various output trace files
that have been used both to do data processing and to visualize
the simulation. Within our simulation we  used the visualization
tool of Network Animator NAM  and the NS-2 trace files analyzer of
Tracegraph. For the simulation of mobility we used the Setdest
program in order to generate movement pattern files using the
random Waypoint algorithm.

An example of simulation is shown graphically in Figure
2. Basically, it consists of generating a scenario file that
describes the movement pattern of the nodes and a communication
file that describes the traffic in the network. These files are
used to produce trace files that are analyzed to measure various
parameters. An excerpt of the trace files corresponding to the
same example is shown in Table \ref{tab:Trace}.

\begin{figure*}[htb]
  \centering
     \includegraphics[scale=0.6]{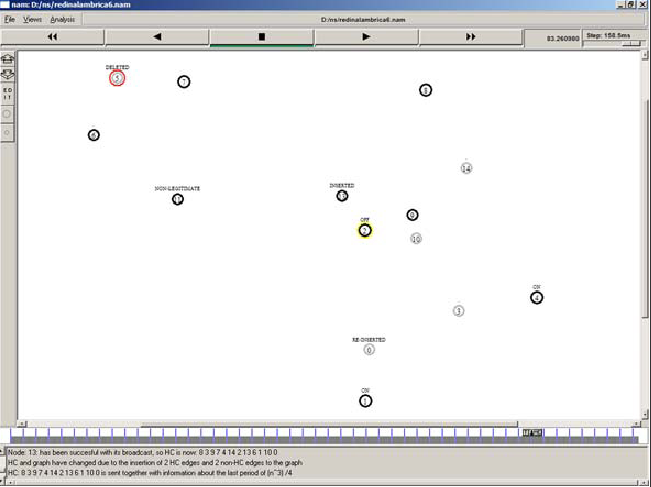}
  \caption{Example of Network Simulation with NS-2}
  \label{fig:NS}
\end{figure*}

\begin{figure*}[htb]
  \centering
     \includegraphics[scale=0.3]{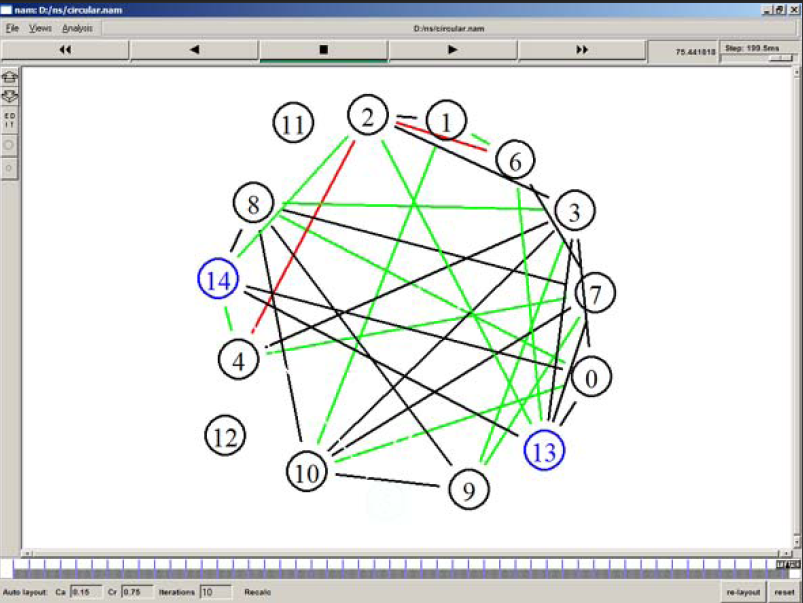}
  \caption{Example of Final Associated Graph and Hamiltonian Cycle} \label{fig:Graph}
  \vspace{-0.2cm}
\end{figure*}

\begin{table*}[htb]
\begin{center}
\begin{tabular}{|c|c|c|} \hline
Time& Event  & HC  \\ \hline

0.1 & \scriptsize{0, 1, 2, 3, 4, 5, 6, 7, 8, 9, 10 are legitimate}  &
\scriptsize{8,3,9,7,4,2,6,5,1,10,0}\\

1.29 & \scriptsize{Insertion of Node 14 is broadcast by Node 4}  &
\scriptsize{8,3,9,7,4,14,2,6,5,1,10,0  } \\

1.30 &\scriptsize{ Nodes 3, 1, 0 do not answer to the proof of life}  &
\\

3.29 & \scriptsize{Node 0 reaches 8 and starts a ZKP for re-insertion} & \\

8.69 & \scriptsize{Node 3 reaches 4 and starts a ZKP for re-insertion } & \\

9.40 &\scriptsize{ Node 1 reaches 10 and starts a ZKP for re-insertion} &
\\

11.65 &\scriptsize{ Node 1 turns off }& \\

13.97 & \scriptsize{Proof of life started by Node 3} & \\

14.27 & \scriptsize{Nodes 1, 2 do not answer to the proof of life}  & \\

14.82 &\scriptsize{ Node 2 reaches 14 and starts a ZKP for re-insertion } &
\\

17.27 &\scriptsize{ Proof of life started by Node 2} & \\

17.57 & \scriptsize{Nodes 1, 5 do not answer to the proof of life  } & \\

21.71 & \scriptsize{Node 5 turns off }& \\

31.40 & \scriptsize{Node 1  turns on and Node 2 is chosen for the ZKP }& \\

31.46 &\scriptsize{ Node 4 turns off }& \\

32.51 & \scriptsize{Proof of life started by Node 1} & \\

32.78 & \scriptsize{Nodes 4, 5, 6 do not answer to the proof of life}  & \\

34.29 &\scriptsize{ Node 6 reaches 2 and starts a ZKP for re-insertion } & \\

38.51 & \scriptsize{Proof of life started by Node 6} & \\

38.79 & \scriptsize{Nodes 4, 5 do not answer to the proof of life} & \\

41.46 & \scriptsize{Node 1 turns off} & \\

53.25 & \scriptsize{Node 1  turns on and Node 0 is chosen for the ZKP }& \\

59.61 &\scriptsize{ Proof of life started by Node 6 }& \\

59.99 & \scriptsize{Nodes 4, 5 do not answer to the proof of life }& \\

64.26 & \scriptsize{ Node 5 is deleted} &
\scriptsize{8,3,9,7,4,14,2,6,1,10,0}\\

64.71 & \scriptsize{Node 2 turns off} & \\

72.58 & \scriptsize{Node 4  turns on and Node 0 is chosen for the ZKP}  & \\

75.41 & \scriptsize{Insertion of Node 13 is broadcast by Node 14 } & \
\scriptsize{8,3,9,7,4,14,2,13,6,1,10,0 }  \\

75.43 & \scriptsize{Node 2 does not answer to the proof of life } &
\\

 \hline
\end{tabular}
\caption{Example of Trace} \label{tab:Trace}
\end{center}
 
\end{table*}

The trace files are used to visualize the simulation using NAM,
while the measurement values are used as data for plots with
Tracegraph. The final graph and Hamiltonian cycle associated to
the example network is shown in Figure 3
 where green is used to indicate the Hamiltonian cycle, blue is used for
the inserted nodes and red is used for the
 edges deleted from the Hamiltonian cycle when inserting new
 nodes.

In order to study the effectiveness of the GASMAN, we studied it in a set of realistic scenarios. In particular, we used the most commonly used mobility model by the research community, the so-called Random Waypoint Model, which uses pause times and random changes in destination and speed.

An extensive number of simulations using NS-2 simulator with 802.11 MAC and DSR
routing protocols in order to see the effects of different metrics by varying network density and topology were run. Within the simulations, relationships can be established anytime two nodes are located in close proximity and the random walk mobility model was used with various pause time and maximum speed. In particular, we varied the number of nodes from 15 to 100. Also, our architecture was evaluated with 250 x 250, 500 x 500, and 750 x 750 m2 square area of ad-hoc network. In each case, the nodes move around with 0.5
second pause time and 20m/s maximum speeds. The transmission range of the
secure channel is 5 meters while that of the data channel is fixed to 250 meters.
The period of simulation varied from 60 to 200 seconds. We also changed the
probabilities of insertions and deletions in each second from 5\%
to 25\%, in order to modify the mobility rate and antenna range of
nodes from 2 to 15 m/s and 100 to 250 meters respectively. This
range also defines different frequencies of accesses to the
network.

The first conclusions we obtained from the simulations
are:
\begin{itemize}
\item The  protocol  scales perfectly to any sort of networks with different levels of topology changes.
\item Node density is a key factor for the mean time of insertions, but such a factor is not as big as it might be
 previously assumed since nodes do not forward two packets of data corresponding to the same proof of life
 coming from two different nodes.
\item A right choice of parameter $T$ should be done according to number of nodes, bandwidth of wireless connections and
computation and storing capacities of nodes.
\item  A positive aspect of the proposal is that the requirements in the devices' hardware are very low.
\end{itemize}

\section{Conclusions and Open Questions}

Successful authentication in mobile ad-hoc networks is critical for assuring secure and effective operation of the supported application. This work describes a new authentication scheme, the so-called GASMAN, which was
specially designed for MANETs. Such a protocol supports
knowledge-based member authentication  in server-less
environments. The overall goal of the GASMAN has been to design
a strong authentication scheme that is able to react and adapt to
network topology changes without the necessity of any centralized
authority. Its core technique consists of a Zero-Knowledge Proof,
in order to avoid the transference of any relevant information.
Furthermore, the proposal is balanced since the procedures that
the legitimate members of the network have to carry out when the
network is updated (insertion or deletion of nodes) imply
identical work for every legitimate member of the network.

The development of an initial simulation of the proposal through
the NS-2 network simulator has been carried out. The definitive
simulation results will be included in a forthcoming version of
this work. Also, the study of different applications, practical
limitations and possible extensions of the GASMAN may be
considered open problems.

\end{document}